\documentclass[11pt,eclepsf]{article} 
\usepackage{graphicx}
\usepackage{amssymb}
\usepackage{amsmath}

\newcommand{\beq}{\begin{eqnarray}}
\newcommand{\eeq}{\end{eqnarray}}

\newcommand{\del}{\partial}
\def\gsim{\displaystyle\mathop{>}_{\sim}}
\def\lsim{\displaystyle\mathop{<}_{\sim}}

\makeatletter  
  
\@addtoreset{equation}{section} 
\makeatother 
\def\Slash#1{/\hspace{-0.23cm}{#1}} 
    \def\ket{\rangle}
\def\del{\partial}
\def\ms{M_S} \def\mq{m_q} \def\ma{M_A}
\begin{document} 
\title{Two Nucleon-States in a Chiral Quark-Diquark Model} 
\author{Keitaro Nagata and Atsushi Hosaka,\\ 
{\it Research Center for Nuclear Physics,  Osaka University}\\
{\it Ibaraki 567-0047, Japan}}
\date{}
\maketitle 
\begin{abstract}
We study the ground and first excited states of nucleons 
in a chiral quark-diquark model.  
We include two quark-diquark channels of the scalar-isoscalar
and axial-vector-isovector types for the nucleon states.    
The diquark correlation violating the spin-flavor SU(4)$_{SF}$ symmetry allows
to treat the two quark-diquark channels independently. Hence the two
states appear as the superpositions of the two quark-diquark channels;
one is the nucleon and the other is a state which does not appear in
the SU(4)$_{SF}$ quark models.
With a reasonable choice of model parameters, the mass 
of the excited state appears at around 1.5 GeV, 
which we identify with the Roper resonance $N(1440)$. 
\end{abstract} 

\section{Introduction}
Hadron spectrum should in principle be understood by 
Quantum Chromodynamics (QCD).  
However, due to the difficulties of non-perturbative nature of 
QCD, in practice, QCD oriented models are often used 
for the description of hadrons.  
The non-relativistic quark model is one of successful ones\cite{DeRujula:1975ge,Isgur:1978xj,Isgur:1979be,hosaka_book}.  
Employing a confining potential which is usually taken as
a harmonic oscillator, various hadrons are described as 
single particle states of two (for mesons) and three (for baryons)
quark states.  
Being not always said explicitly, the SU(4)$_{SF}$ spin-(two) flavor 
symmetry is usually 
assumed to work well, and possible interactions such 
as spin-color and spin-flavor interactions are treated 
perturbatively.

Such residual interactions may play an essential role 
for hadron properties by determining symmetries of hadrons.  
One of efficient methods which takes into account 
such interactions is to consider diquark correlations.  
In fact, it is known that the spin-color and spin-flavor 
interactions lead to a strongly attractive correlation 
for a color, flavor and spin antisymmetric quark-quark 
pair (scalar diquark)~\cite{Jaffe:1976ig,Jaffe:2003sg}.
Contrary, the color symmetric states receive repulsive correlations.  
If such correlations are strong enough, the resulting 
hadron properties will be quite different from what 
are expected from the SU(4)$_{SF}$ quark models.  

In this paper, we would like to consider the case 
in which diquarks receive sufficiently strong correlation 
such that they are treated as independent 
degrees of freedom.  
This is the SU(2)$_L\times$SU(2)$_R$ chiral quark-diquark model which
we studied
previously\cite{Ebert:1997hr,Abu-Raddad:2002pw,Nagata:2003gg,Nagata:2004ky}.  

It was shown that there are two low lying diquarks with color
$\bar{3}$, one is the scalar and isoscalar ($\equiv$ scalar) diquark, 
and the other is the axial-vector and isovector ($\equiv$ axial-vector)
diquark\cite{Espriu:1983hu}.
In a quark model picture, they are both in the 
ground state but with different spin and isospin configuration;   
the former is spin-singlet and flavor-singlet and the latter is spin-triplet 
and flavor-triplet ones.  
Adding another quark to these diquarks we can construct two
states having the nucleon quantum numbers $J^P = 1/2^+$, i.e. $\left[
[1/2, 1/2]^0, 1/2 \right]^{1/2}$ and $\left[
[1/2, 1/2]^1, 1/2 \right]^{1/2}$.  
In the SU(4)$_{SF}$ quark models, such two states do not 
exist independently, but only their linear combination is realized
(the sum of the $\rho$ and $\lambda$ symmetric
states)~\cite{hosaka_book,close_book}. If the diquark correlation exists, these two states may be treated independently.  


Having the two states independently in the model set up, the two 
nucleon states emerge naturally.  
It is known that the scalar diquark is lighter than the axial-vector 
diquark, whose mass difference may be related to the 
mass difference between the nucleon and delta, which is of order 300 MeV.  
Hence we have the two nucleon states with the mass difference 
of this order to start with.  
Introducing a mixing interaction between them, 
the two states repell each other.  
One would then expect additional, but not too large,  
mass splitting due to the mixing interaction.   
Altogether, we expect a mass difference about 4-500 MeV. 
A nucleon excited state with an excitation energy of this
amount is naturally identified with the Roper resonance $N(1440)$.  
The Roper state appears as a spin partner of the nucleon in the sense that 
it has a diquark component of different spin.  
Hence the mass splitting is generated by residual (hyper-fine) 
interactions and is likely not too large.  
This picture is very much different from the conventional 
picture of the Roper resonance of the excitation of 
$2\hbar \omega$ of the harmonic oscillator potential for 
confinement\cite{Glozman:1995fu,Hosaka:1999aa} or breathing mode of
the bag model or Skrymion\cite{Brown:1983ib,Hatsuda:1986ph,Breit:1984tf,Hayashi:1984bc,Zahed:1984qv,Mattis:1984ak,Hosaka:1988pj}.
In this paper, we demonstrate such a realization of the Roper 
resonance as a partner of the ground state of the nucleon 
in the quark-diquark model with a path-integral hadronization method.  

This paper is organized as follows. 
In \S\ref{sec:model} we introduce the quark-diquark model with scalar
and axial-vector diquarks. Chiral symmetry of the model is briefly
discussed in the nonlinear representation.  
In \S\ref{sec:hadronize} the model is hadronized to obtain an
effective meson-baryon Lagrangian in the path-integral method. 
The obtained Hamiltonian is diagonalized in the two dimensional 
space of the two nucleons constructed by the scalar and axial-vector
channels.  
In \S\ref{sec:numerics} numerical results are presented and the 
higher nucleon state is identified with the Roper resonance. 
The final section is devoted to a summary.

\section{Framework}
\label{sec:model}
We start from the SU(2)$_L\times$ SU(2)$_R$ chiral quark-diquark model of two diquarks~\cite{Abu-Raddad:2002pw,Nagata:2004ky}, 
\begin{eqnarray}
{\cal L} = \bar{\chi}_c(i\rlap/\del - m_q) \chi_c \;+\;
D^\dag_c (\del^2 + M_S^2)D_c
 +
{\vec{D}^{\dag\;\mu}}_c 
\left[  (\del^2 + M_A^2)g_{\mu \nu} - \del_\mu \del_\nu\right]
\vec{D}^{\nu}_c +{\cal L}_{int},
\label{lsemibos}
\end{eqnarray}
where $\chi_c$, $D_c$ and $\vec{D}_{\mu c}$ are the constituent quark, scalar
diquark and axial-vector diquark fields, and $m_q$, $M_S$ and $M_A$
are the masses of them. The indices $c$ represent the color. Note that the 
diquarks microscopically correspond to the bi-linears of two
quarks\cite{Abu-Raddad:2002pw}; $D_c\sim
\epsilon_{abc}\tilde{\chi}^b\chi^c,\ \vec{D}_{\mu c}\sim \epsilon_{abc}\tilde{\chi}^b \gamma_\mu \gamma_5 \vec{\tau}\chi^c$.
Here $\tilde{\chi}=\chi^T C\gamma_5 i\tau_2$ and $\epsilon_{abc}$ is
the antisymmetric tensor, then both the diquarks belong to color anti-triplet.
The term ${\cal L}_{int}$ is the quark-diquark interaction, which is written as
\begin{equation}
{\cal L}_{int}=G_S\bar{\chi}_cD^\dagger_c
D_{c^\prime}\chi_{c^\prime}+v(\bar{\chi}_cD^\dagger_c\gamma^\mu\gamma^5
\vec{\tau}\cdot\vec{D}_{\mu c^\prime} \chi_{c^\prime}+\bar{\chi}_c\gamma^\mu\gamma^5
\vec{\tau}\cdot\vec{D}^\dagger_{\mu c}
D_{c^\prime}\chi_{c^\prime})+G_A\bar{\chi}_c\gamma^\mu\gamma^5
\vec{\tau}\cdot\vec{D}^\dagger_{\mu c}
\vec{\tau}\cdot\vec{D}_{\nu c^\prime}\gamma^\nu\gamma^5 \chi_{c^\prime},
\label{eq:twoc}
\end{equation}
where $G_S$ and $G_A$ are the coupling constants for the quark and scalar
diquark, and for the quark and  axial-vector diquark. The coupling
constant $v$ causes the mixing between the scalar and axial-vector
channels in the nucleon wave-functions. 

Since chiral symmetry is important for hadron physics, 
we briefly discuss the properties of our model Lagrangian 
(\ref{lsemibos}) under chiral transformation.  
In our formulation we employ the non-linear representation of chiral
symmetry\cite{Abu-Raddad:2002pw}, therefore the constituent quarks are
transformed as
\begin{equation}
\chi_c\to h(x)\chi_c .
\end{equation}
Here $h(x)$ is a local transformation 
depending on the global SU(2)$_L\times$ SU(2)$_R$ transformation
 and the pion field at $x$. Then, the baryon field written as a product of 
a scalar diquark and a quark, and of an axial-vector diquark and a
quark are transformed as (in detail see Appendix.~\ref{sec:app1})
\begin{subequations}
\begin{eqnarray}
D_c\chi_c&\to& h(x) D_c\chi_c,\\
\vec{D}_{\mu c}\cdot\vec{\tau}\gamma^\mu\gamma^5\chi_c&\to& h(x) \vec{D}_{\mu c}\cdot\vec{\tau}\gamma^\mu\gamma^5\chi_c.
\end{eqnarray}
\label{eq:transformations}
\end{subequations}
Note that the both baryon operators are transformed in the same way as the
quark is under the chiral SU(2)$_L\times$ SU(2)$_R$ transformation.  
In the non-linear representation, the kinetic term of the 
quark in Eq.~(\ref{lsemibos}) contains the mesonic
currents~\cite{hosaka_book}. 
However, we do not show them, because they are not important in the
present discussions.

\section{Hadronization and the self-energies of nucleons}
\label{sec:hadronize}
\begin{figure}[tbh]                                                       
\centering{                                                               
\includegraphics[width=5cm]{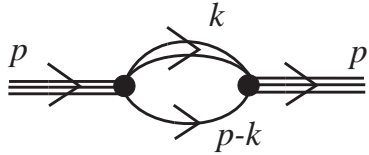}
\begin{minipage}{14cm}                                                    
   \caption{\small A diagrammatic representation of the quark-diquark     
   self-energy. The single, double and triple lines represent the quark,  
   diquark and nucleon respectively. The blobs represent the three point  
   quark-diquark-baryon interactions.}                                    
   \label{fig:self}                                                       
 \end{minipage}}                                                          
\end{figure}

To perform the hadronization procedure, we introduce the
auxiliary fields for baryons
\begin{eqnarray}
{\cal L}= \bar{\chi}(i\rlap/\del - m_q) \chi \;+\;
D^\dag (\del^2 + M_S^2)D
 +
{\vec{D}^{\dag\;\mu}} 
\left[  (\del^2 + M_A^2)g_{\mu \nu} - \del_\mu \del_\nu\right]
\vec{D}^{\nu}+\bar{\psi}\hat{G}\psi-\bar{B}\hat{G}^{-1}B.
\label{L_inter}
\end{eqnarray}
From here, we omit the color indices for brevity.
$B=(B_1, B_2)^T$ is a two component auxiliary baryon field, whose components 
correspond to  scalar and axial-vector channels; $B_1\sim D\chi$ 
and $B_2\sim \vec{\tau}\cdot\vec{D}_\mu\gamma^\mu\gamma^5\chi$.
In Eq.~(\ref{L_inter}) we have introduced matrix notations as
\begin{eqnarray}
\psi&=&\left(\begin{array}{c} D\chi\\ 
\vec{D}_\mu\cdot\vec{\tau}\gamma^\mu\gamma^5\chi\end{array}\right),\quad
\bar{\psi}=
\left(\begin{array}{cc} 
\bar{\chi}D^\dagger, & 
\bar{\chi}\vec{D}_\mu^\dagger\cdot\vec{\tau} \gamma^\mu \gamma^5 \end{array} \right),\\
\hat{G}&=&\left(\begin{array}{cc} G_S & v \\ v & G_A \end{array}\right).
\end{eqnarray}
Through the hadronization
procedure\cite{Abu-Raddad:2002pw,Cahill:1988zi,Reinhardt:1989rw}, the
quark and diquark fields are eliminated and a meson-baryon Lagrangian
in the $\mbox{tr}\log$ form is obtained as
\begin{equation}
{\cal L}=-\bar{B}\hat{G}^{-1}B+i\mbox{Tr}\ln(1-A).
\label{eq:trln}
\end{equation}
Here the matrix $A$ is defined by
\begin{subequations}
\begin{eqnarray}
A&=&\left(\begin{array}{cc} a_{11} & a_{12}\\ a_{21}& a_{22}\end{array}\right),\\
a_{11}&=&\Delta^T \bar{B}_1 S B_1,\\
a_{12}&=&\Delta^T\bar{B}_2 \tau^i \gamma^\nu\gamma^5  S B_1,\\
a_{21}&=&(\Delta_{\rho\nu}^{lj})^T \bar{B}_1 S \gamma^\mu\gamma^5\tau^j  B_2,\\
a_{22}&=&(\Delta_{\rho\nu}^{lj})^T \bar{B}_2\gamma^\nu\gamma^5\tau^i S \gamma^\mu\gamma^5\tau^j B_2.
\end{eqnarray}
\end{subequations}
where $S$, $\Delta$ and $\Delta_{\mu\nu}$ are the propagators of the
quark, scalar diquark and axial-vector diquark, respectively.
The expansion of the tr log in Eq.~(\ref{eq:trln}) gives various
terms. 
The first term of the expansion gives the self-energies of the 
nucleons as
\begin{eqnarray}
{\cal L}=
\bar{B}\left(\begin{array}{cc} 
\Sigma_S(p) & 0 \\ 0 & \Sigma_A(p)\end{array}\right)B 
- \frac{1}{|\hat{G}|}\bar{B}\left(\begin{array}{cc} G_A 
& -v \\ -v & G_S \end{array}\right) B,
\label{eq:lag2}
\end{eqnarray}
where $|\hat{G}|=\mbox{det} \hat{G}=G_SG_A-v^2$.
The scalar and axial-vector diquark contributions to the
self-energies, $\Sigma_S$ and $\Sigma_A$, 
are shown diagrammatically in Fig.~\ref{fig:self} and are computed as
\begin{subequations}
\begin{eqnarray}
\Sigma_S(p)&=&-i N_c\int\frac{d^4k}{(2\pi)^4}
\frac{1}{k^2-\ms^2}\frac{\Slash{p}-\Slash{k}+\mq}{(p-k)^2-\mq^2},\\
\Sigma_A(p)&=&-iN_c\int\frac{d^4k}{(2\pi)^4}
\frac{k^\mu k^\nu/\ma^2-g^{\mu\nu}}{k^2-\ma^2}\delta_{ij}\gamma_\nu\gamma_5\tau_j
\frac{\Slash{p}-\Slash{k}+\mq}{(p-k)^2-\mq^2}\tau_i\gamma_\mu\gamma_5 . 
\end{eqnarray}
\label{eqn:selfa}
\end{subequations}
Here $N_c$ is the number of colors. 
Since the self-energies Eqs.~(\ref{eqn:selfa}) are divergent, 
we regularize them 
by the three momentum cutoff scheme\cite{Nagata:2004ky}.

The self-energies $\Sigma_S$ and $\Sigma_A$ may be
decomposed into the scalar and vector parts as
\begin{subequations}
\begin{eqnarray}
\Sigma_S(p_0)-\frac{1}{|\hat{G}|}G_A&=&Z_S^{-1}(p_0\gamma^0-a_S),
\label{SigmaS}\\
\Sigma_A(p_0)-\frac{1}{|\hat{G}|}G_S&=&Z_A^{-1}(p_0\gamma^0-a_A),
\label{SigmaA}
\end{eqnarray}
\label{Sigma}
\end{subequations}
where we employ the nucleon rest frame $p_\mu=(p_0,\ \vec{0})$.
The bare baryon fields $B_{1,2}$ are now renormalized as
\begin{eqnarray}
\left(\begin{array}{c} B_1 \\ B_2 \end{array}\right)
=\left(\begin{array}{c}\sqrt{Z_S} B_1^\prime\\
\sqrt{Z_A} B_2^\prime\end{array}\right) ,
\end{eqnarray}
with which the Lagrangian (\ref{eq:lag2}) is reduced to
\begin{equation}
{\cal L}=\bar{B}^\prime(p_0\gamma^0 - \hat M )B^\prime\, ,
\end{equation}
where the mass matrix $\hat{M}$ is given as
\begin{eqnarray}
\hat{M}=\left(\begin{array}{cc} a_S & -\sqrt{Z_S Z_A}
\frac{v}{|\hat{G}|}\\ -\sqrt{Z_S Z_A} \frac{v}{|\hat{G}|} & a_A
\end{array}\right).
\label{eq:mmatrix}
\end{eqnarray}
When there is no mixing interaction $(v=0)$, $B_{1,2}^\prime$ become
the physical baryon fields and $a_{S,A}$  the physical masses.
On the contrary,
in the presence of the mixing, the physical states are obtained
after the diagonalization of the mass matrix 
by an unitary transformation:
\begin{eqnarray}
B^\prime&=&U^\dagger N,\\
U\hat{M}U^\dagger&=&\left(\begin{array}{cc} M_1 & 0 \\ 0 & M_2 \end{array}\right).
\end{eqnarray}
One finds
\begin{equation}
{\cal L}=\bar{N_1}(p_0\gamma^0-M_1)N_1+\bar{N_2}(p_0\gamma^0-M_2)N_2,
\end{equation}
where the physical eigenvalues $M_{1,2}$ and eigenvectors 
$N=(N_1,\ N_2)^T$ are obtained as
\begin{eqnarray}
M_{1,2}&=&\frac12\left[a_S+a_A\pm
\sqrt{(a_S-a_A)^2+4Z_S Z_A
\left(\frac{v}{|\hat{G}|}\right)^2}\right]\, , \\
N_1&=&\cos\phi B_1^\prime+\sin\phi B_2^\prime\\
N_2&=&-\sin\phi B_1^\prime+\cos\phi B_2^\prime \, , 
\end{eqnarray}
and the mixing angle $\phi$ is given by 
\begin{equation}
\tan 2\phi=\frac{2\sqrt{Z_S Z_A}v}{(a_A-a_S)|\hat{G}|}
\end{equation}

%
\section{Numerical results}
\label{sec:numerics}
For numerical calculations, let us first discuss model parameters. 
The constituent mass of the $ud$ quarks $m_q$ and the three momentum 
cutoff $\Lambda$ are determined in such a way that they 
reproduce meson properties in the NJL model\cite{Vogl:1991qt,Hatsuda:1994pi}. 
The masses of the diquarks may be also calculated in the NJL
model~\cite{Vogl:1991qt,Cahill:1987qr}. Here we choose $m_q$=390 MeV,
$M_S$=650 MeV, $M_A$=1050 MeV and $\Lambda$=600 MeV, 
which are within the reasonable range known from 
the previous study of diquarks in the NJL
model. The mass difference $M_A - M_S$
may be related to that of the nucleon and delta.  
In the quark-diquark models, the delta is expressed as a bound state 
of an axial-vector diquark and a quark, while the nucleon is a 
superposition of the two components:
\beq
| \Delta \ket &\sim& \vec D_\mu \chi \, ,
\nonumber \\
|N \ket &\sim& \cos \theta D \chi + \sin \theta 
\vec \tau \cdot \vec D_\mu \gamma^\mu \gamma_5 \chi \, .\nonumber
\eeq
Therefore, in a simple additive picture, the $N-\Delta$ mass difference 
can be expressed as
\beq
M_\Delta - M_N = \cos^2 \theta (M_A - M_S) \, . 
\label{massdiff}
\eeq
In deriving this relation we have ignored possible interactions 
such as the binding effect of the quarks and diquarks, and 
the interaction between the two diquark channels.  
However, for a rough estimation we may use the 
relation (\ref{massdiff}) and the mixing angle 
$\cos^2 \theta \gsim 1/2$, assuming that the nucleon 
state is rather dominated by the scalar diquark component.  
This qualitatively justifies the mass difference 
$M_A - M_S \sim 400$ MeV that we adopt.  
The masses of the two nucleon states are then studied 
in the following two cases.  
%

\begin{figure}[h]
\centering
\includegraphics[width=6cm]{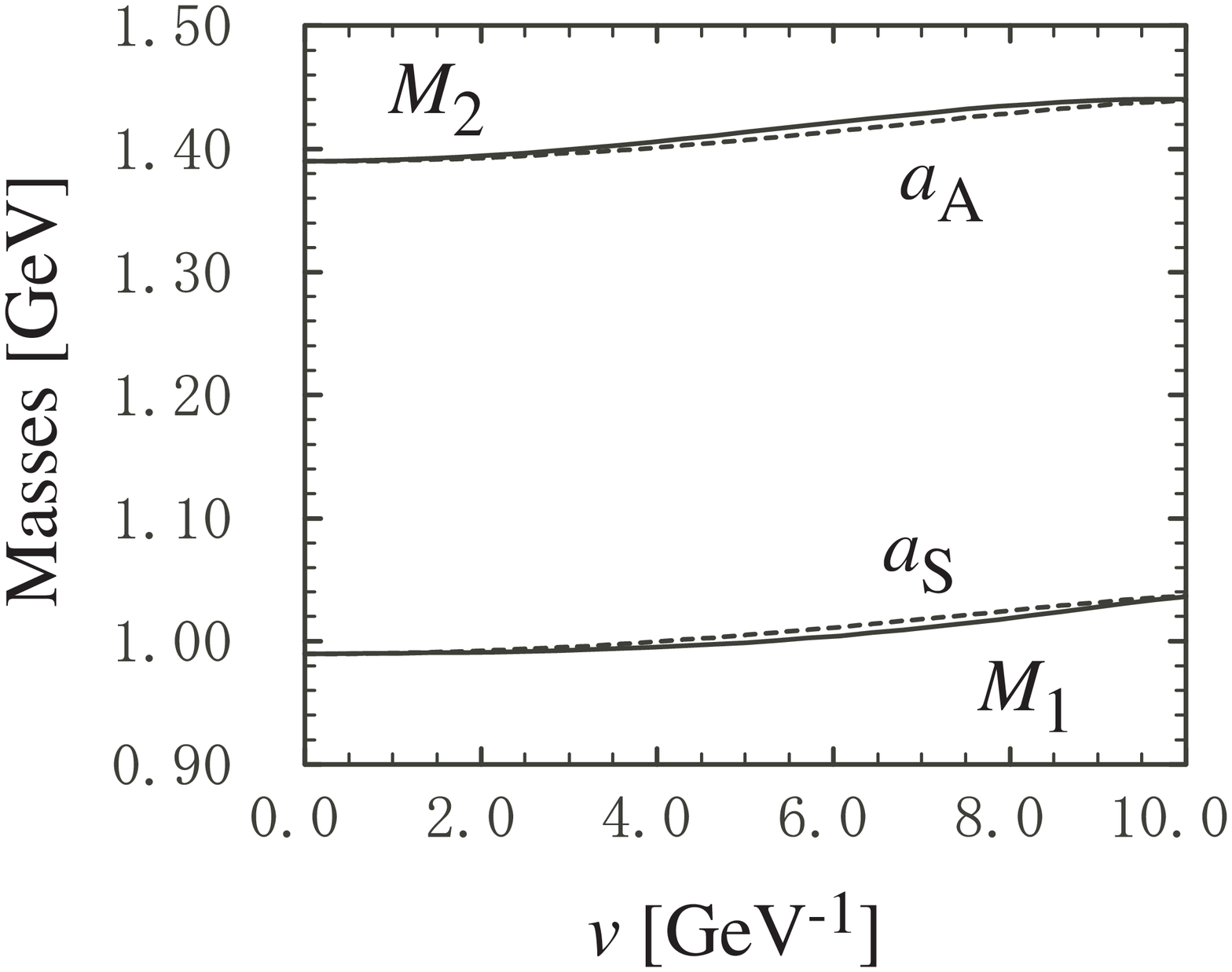}
\includegraphics[width=6cm]{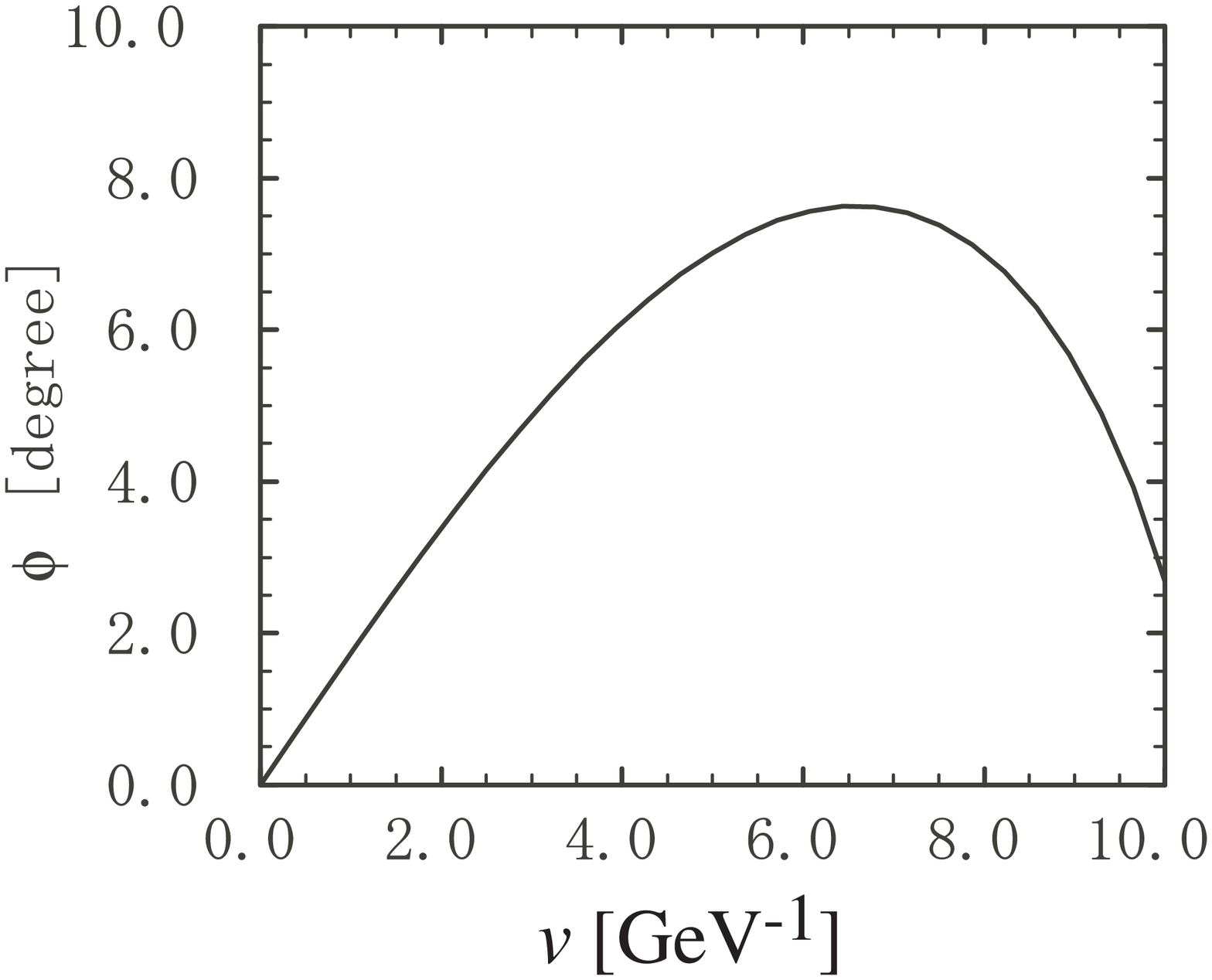}
\caption{The mixing interaction $v$ dependence of the masses $M_{1,2}$ (left)
and the mixing angle $\phi$ (right) with the fixed values of the
quark-diquark coupling constants $G_{S,A}$. In the left panel the
solid lines represent $M_1$ and $M_2$ and the dashed lines represent
$a_S$ and $a_A$. In this plot, the coupling constants $G_S$ and $G_A$
are fixed as $G_S$=54 GeV$^{-1}$ and $G_A$=5.9 GeV$^{-1}$ for all range
of the values of $v$.}\label{fig:mass1}
\end{figure}

(i) In the first case, 
$G_S$ and $G_A$ are fixed such that the binding energies 
of the two quark-diquark bound states become 50 MeV when there is 
no coupling $v$. The resulting coupling strengths are 
$G_S \sim 54$ and $G_A \sim 5.9$, which generates the masses $M_1=0.99$
GeV and $M_2=$1.39 GeV at $v=$ 0 GeV$^{-1}$. In the
previous works\cite{Abu-Raddad:2002pw,Nagata:2003gg}, this binding
energy was chosen in order to obtain a reasonable size of the
nucleon. The masses and the mixing angle $\phi$ of the two nucleon states are 
then calculated as functions of the coupling strength $v$, which are
shown in Figs.~\ref{fig:mass1}. The effect of the $v$-coupling appears not only
in the off-diagonal elements but also in the diagonal elements of the
mass matrix (\ref{eq:mmatrix}). The effect of the $v$-coupling on the diagonal
elements is from the term of $|\hat{G}|$ in Eqs.~(\ref{Sigma}), which is shown by dashed lines of Fig.~\ref{fig:mass1}. 
As is shown in the figure, the $v$-coupling acts repulsively to both the
diagonal elements $a_{S,A}$. The non-zero off-diagonal elements
then split $M_1$ and $M_2$ as is shown by  the solid lines of
Fig.~\ref{fig:mass1}. 
The off-diagonal coupling decreases the mass of the nucleon and
increases that of the heavier state, while both the two diagonal
elements increase as $v$ is increased. This is why these two
contributions cancel each other for the nucleon, but they are
enhanced for the heavier state. 
We find that $M_1 = 0.99 $ GeV and 
$M_2 =$ 1.44 GeV when $v\sim  9$ GeV$^{-1}$.  
The mass of the second nucleon is close to that of the Roper
resonance $N$(1440). At this strength of $v$, 
as is shown in the right panel of Fig.~\ref{fig:mass1}, 
the mixing angle is rather small, 
$\phi \lsim 10$ degree.  
Even at the small mixing angle, the effect of the 
axial-vector component is significant, since the self-energy 
$\Sigma_A$ is much larger than $\Sigma_S$~\cite{Nagata:2004ky}.  

\begin{figure}[h]
\centering
\includegraphics[width=6cm]{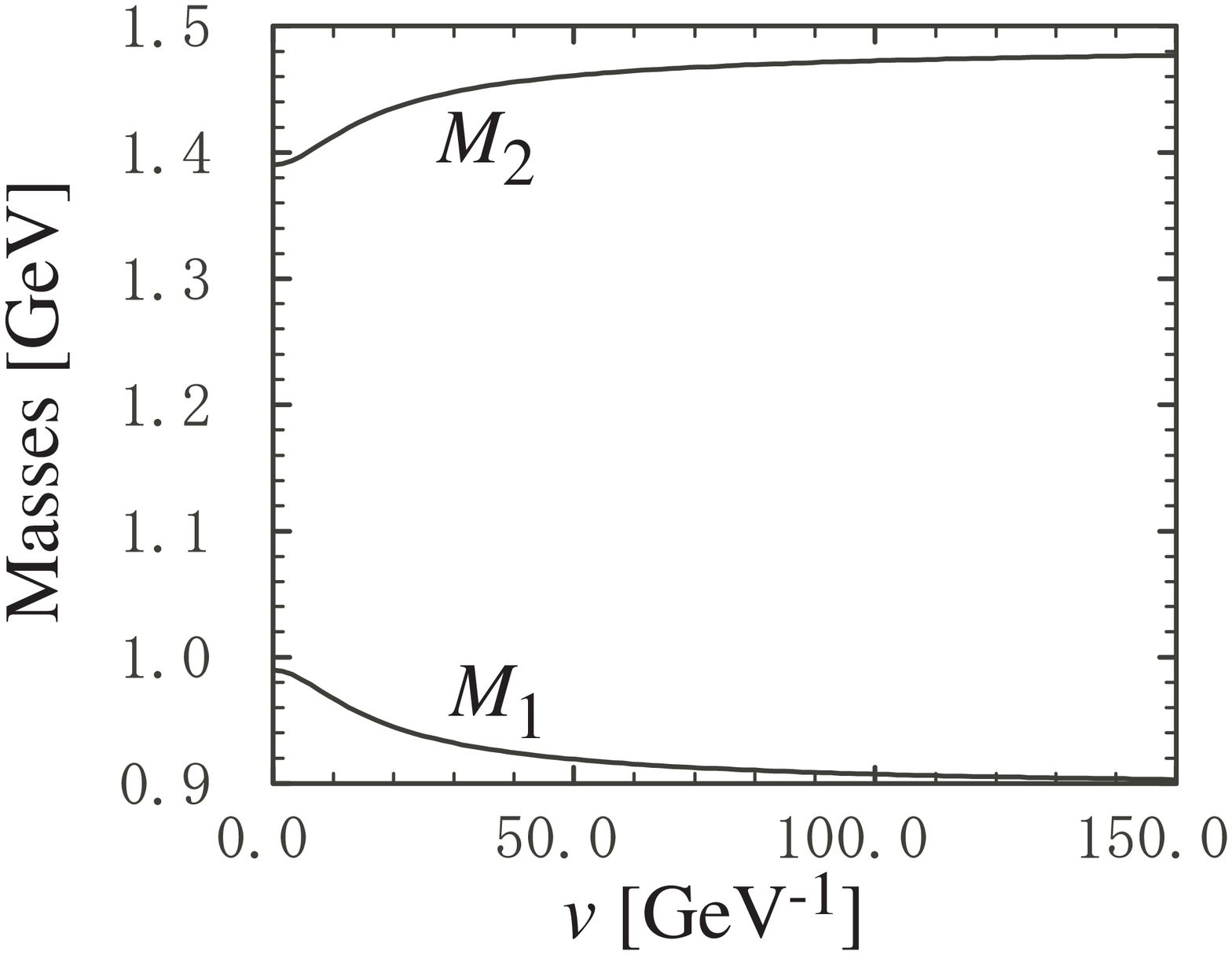}
\includegraphics[width=6cm]{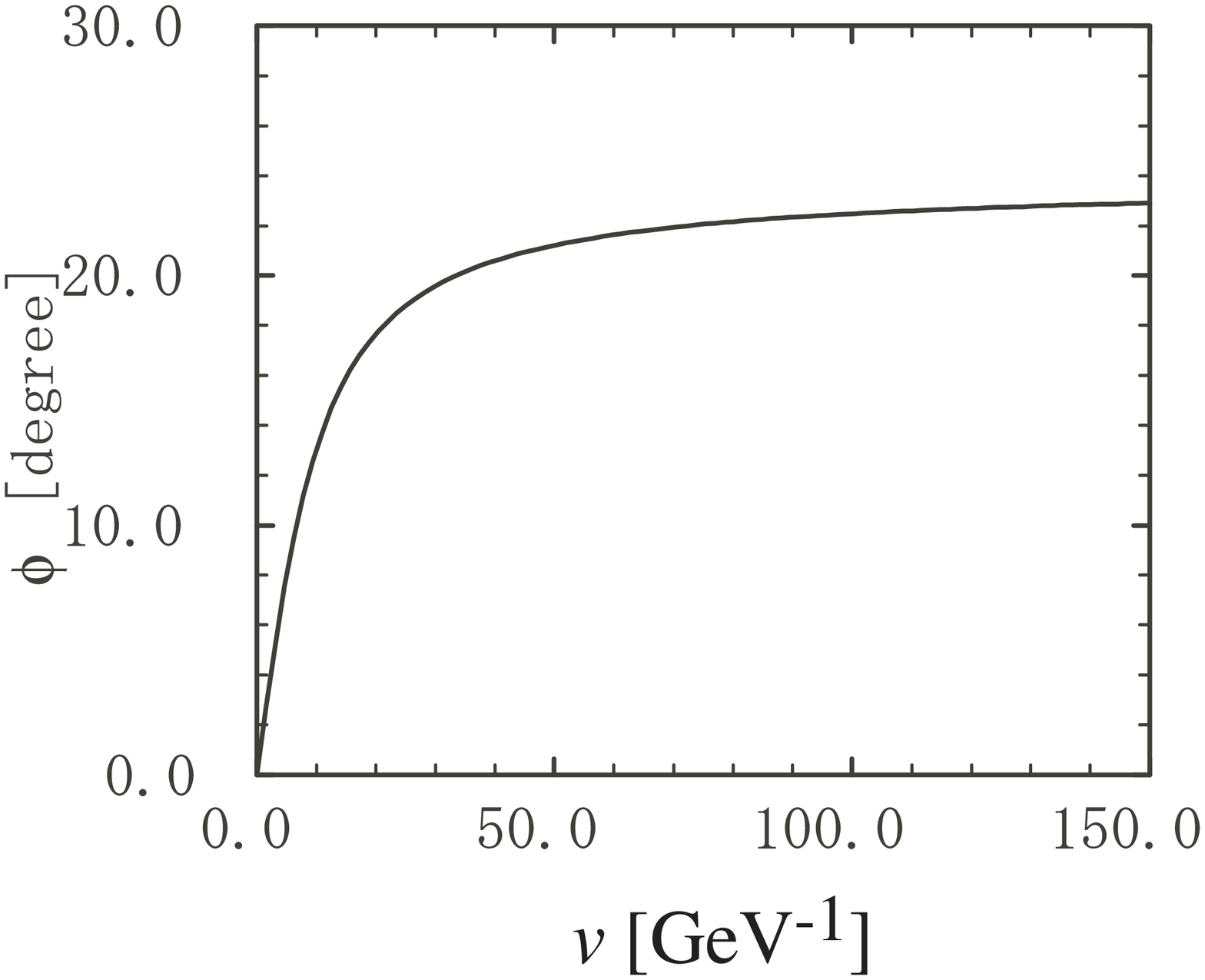}
\caption{The mixing interaction $v$ dependence of the masses $M_{1,2}$ (left)
and the mixing angle $\phi$ (right) with the fixed values of $a_{S,A}$. The
solid lines represent $M_1$ and $M_2$  in the left panel. In this plot
the coupling constants $G_S$ and $G_A$ are determined for each values
of $v$ such that the masses $a_S$ and $a_A$ are generated with the
binding energies of 50 MeV.}\label{fig:mass2}
\end{figure}

(ii) As have been mentioned, the masses $M_{1,2}$ depend not only on 
the off-diagonal elements of the mass matrix (\ref{eq:mmatrix}), but also  
on the diagonal elements.
In the second choice we determine the 
parameters $G_S$ and $G_A$ such that they always produce the 
same amount of the binding energy of 50 MeV as the coupling 
strength $v$ is varied, or $a_S$ and $a_A$ are fixed at 
$a_{S,A}=m_q+M_{S,A}-50$ MeV. In this way we can study the effect of
the off-diagonal coupling $v$ just as in a simple two level problem with
fixed values of the diagonal elements, which is shown in
Fig.\ref{fig:mass2}. In this case dependence of $M_{1,2}$ on the
values of $v$ is larger than that of the case (i).
We find that $M_1=$ 0.94 GeV, $M_2$=1.44 GeV and $\phi$ =18 degree at
$v\sim$ 22 GeV$^{-1}$.

In both the cases of (i) and (ii), we can obtain the reasonable mass $M_2$=1.4
$\sim$1.5 GeV. Hence we identify the second state with the Roper
resonance.


The present identification of the Roper resonance is 
very much different from the conventional picture; 
in the quark model, it is described as an excited 
state of $2\hbar \omega$ with 
$(n,l) = (1,0)$, where $(n,l)$ are the principle and 
angular momentum quantum numbers of the harmonic oscillator.  
The excitation energy of such a state is as high as 
$2 \hbar \omega \sim 1$ GeV for the oscillator parameter 
$\omega \sim 0.5$ GeV, and many mechanisms have been 
proposed to lower the energy~\cite{Glozman:1995fu,Hosaka:1999aa}.  
In the present picture the two nucleons are 
described as quark-diquark bound states, but 
with different diquarks with different spins, isospins and masses.  
In a quark model picture, the two diquarks 
correspond to the quark pair of the 
$\rho\rho$ and $\lambda\lambda$ mixed symmetry
(see Appendix B).  
In the limit of SU(4)$_{SF}$ symmetry, 
these state can not be independent degrees of freedom due to the Pauli
principle when constructing the nucleon state of $J^P = 1/2^+$, 
only their symmetric combination is allowed.  
In the present case, because the 
strong correlation between the quarks violates the 
SU(4)$_{SF}$ symmetry, the two states can be independent.  
In the picture of the harmonic oscillator basis, 
the two quarks of the diquarks are in the ground state 
but with being correlated.  
The energy difference is therefore supplied not by the 
difference in the single particle energies, but 
by the residual but significant correlation 
between the quarks.  
In any event, the scale of the energy of such 
correlations is expected to be of order of 
a few to several hundreds MeV.  
Concerning the wave functions,  the two nucleon states 
of the quark-diquark model may be given in terms of quarks as 
\beq
B_1^\prime &=& D \chi \sim \left[ [1/2, 1/2]^0, 1/2 \right]^{1/2} \, , 
\nonumber \\
B_2^\prime &=& D \chi \sim \left[ [1/2, 1/2]^1, 1/2 \right]^{1/2} \,
. \nonumber
\eeq
From these structure, we may say that 
the Roper resonance appears as a spin partner of the nucleon, which 
has different internal spin structure.  

\section{Summary}
In this paper, we have studied  the nucleon and Roper resonance using
the chiral quark-diquark model. It was shown that in the chiral
symmetric construction of the model, there appear two states as the physical
states with the quantum numbers of the nucleon. It was also shown that
if we identified the low-lying state with the nucleon, the mass of the
higher state became about 1.4 $\sim$ 1.5 GeV, with model parameters
chosen so as to reproduce the mass difference of the nucleon and delta. 
Hence we consider this
explanation of the Roper resonance is natural, although our current
result have some dependence on the model parameters. Although some
authors have studied the Roper resonance by using the lattice QCD\cite{Sasaki:2001nf,Mathur:2003zf,Guadagnoli:2004wm,Burch:2004he,Sasaki:2005ap} the
nature of the Roper resonance is still puzzle. The explanation here is
different from many conventional approach, it may reveal the nature
of the Roper resonance.
 One interesting aspect is the size of the Roper resonance. 
In our picture, the nucleon is expressed as a superposition of the quark
and scalar diquark bound state and the quark and axial-vector diquark
bound state, while the Roper resonance is another superposition
orthogonal to the nucleon state. Then, the difference between the
wave-functions of the quark and scalar diquark bound state and that of
the quark and axial-vector diquark bound state would make that of the
nucleon and Roper resonance different. It is interesting to
study the effect of the change in the wave-functions on various
properties, such as radii, the decay width and so on.

\section*{Acknowledgments}
We thank V. Dmitrasinovic and N. Ishii for fruitful discussions and
useful advice.

\appendix
\section{Chiral transformation of baryon operators}
\label{sec:app1}
Here we explicitly show the transformation of the baryon operators
under the group SU(2)$_L\times$ SU(2)$_R$. The constituent quark field 
is defined through the Weinberg rotation\cite{hosaka_book,Abu-Raddad:2002pw}
\begin{equation}
\chi=\xi_5 q,
\label{eq:wrot}
\end{equation}
where $\xi_5=\exp(i\gamma_5\vec{\tau}\cdot\vec{\Phi}/2f_\pi)$ with the
pion field $\vec{\Phi}$.
Using the right- and left-handed spinors $q_R$ and $q_L$ of linear chiral
representations, Eq.~(\ref{eq:wrot}) is written as
\begin{eqnarray}
\chi=\left(\begin{array}{c} \xi\;q_R\\ \xi^\dagger\;q_L\end{array}\right),
\end{eqnarray}
where $\xi=\exp(i\vec{\tau}\cdot\vec{\Phi}/2f_\pi)$.
Here $\xi$ and $\xi_5$ are related to each other
\begin{subequations}
\begin{eqnarray}
\xi_5+\xi_5^\dagger&=&\xi+\xi^\dagger,\\
\xi_5-\xi_5^\dagger&=&\gamma_5(\xi-\xi^\dagger).
\end{eqnarray}
\end{subequations}
Under the group SU(2)$_L\times$ SU(2)$_R$, $q_R$, $q_L$ and $\xi$ are
transformed as
\begin{eqnarray}
q_{R,L}&\to& g_{R,L}\; q_{R,L},\\
\xi&\to& g_L\xi h^\dagger=h \xi g_R^\dagger,
\end{eqnarray}
then the transformations of the constituent quark field and its Dirac
and charge conjugated fields are given as
\begin{subequations}
\begin{eqnarray}
\chi&\to& h\chi,\\
\bar{\chi}&\to&\bar{\chi}h^\dagger,\\
\tilde{\chi}&\to& \tilde{\chi} h^\dagger.
\end{eqnarray}
\end{subequations}
Then, the two diquark states are transformed as
\begin{subequations}
\begin{eqnarray}
\tilde{\chi}\chi&\to& \tilde{\chi}\chi,\\
\tilde{\chi}\gamma_\mu\gamma_5\tau^a\chi&\to&
R^{ab}(x)\tilde{\chi}\gamma_\mu\gamma_5\tau^b\chi.
\end{eqnarray}
\end{subequations}
Here $R^{ab}(x)$ is the three dimensional rotation matrix, which is
defined by $h^\dagger\tau^a h=R^{ab}(x)\tau^b$. Finally,
the transformations of the baryon operators
Eqs.(\ref{eq:transformations}) are obtained as
\begin{subequations}
\begin{eqnarray}
D\chi&\to& h(x) D\chi,\\
D_\mu^a\gamma^\mu\gamma_5\tau^a\chi&\to& h(x)D_\mu^a\gamma^\mu\gamma_5\tau^a\chi.
\end{eqnarray}
\end{subequations}
\section{Non Relativistic limit}
It is useful to consider the connection to the non-relativistic
quark model.
The interaction Eq.~(\ref{eq:twoc}) may be rewritten as
\begin{eqnarray}
{\cal L}_{int}&=&a\bar{\chi}(\cos\theta
D^\dagger+\frac13\sin\theta\gamma^\mu\gamma^5\vec{\tau}\cdot\vec{D}_\mu^\dagger)(\cos\theta
D+\frac13\sin \theta \vec{\tau}\cdot\vec{D}_\mu\gamma^\mu \gamma^5)\chi\nonumber\\
&+&b\bar{\chi}(-\sin\theta
D^\dagger+\frac13\cos\theta\gamma^\mu\gamma^5\vec{\tau}\cdot\vec{D}_\mu^\dagger)(-\sin\theta
D+\frac13\cos\theta \vec{\tau}\cdot\vec{D}_\mu\gamma^\mu \gamma^5)\chi,
\label{eq:alt}
\end{eqnarray}
with the relations between the parameters
\begin{subequations}
\begin{eqnarray}
a\cos^2\theta+b\sin^2\theta&=&G_S,\\
a\sin^2\theta+b\cos^2\theta&=&9 G_A,\\
(a-b)\sin\theta\cos\theta&=&3v.
\end{eqnarray}
\end{subequations}
Here we note that the baryon fields, in terms of a quark and a diquark,
are related to the nucleon wave-functions of the constituent quark
model by 
\begin{subequations}
\begin{eqnarray}
D\chi=2\phi_\rho\chi_\rho,\\
\vec{D}_\nu\cdot\vec{\tau}\gamma^\nu\gamma_5\chi=6\phi_\lambda\chi_\lambda,
\end{eqnarray} 
\end{subequations}
in the non-relativistic limit, where $\phi_\rho,\ \phi_\lambda$ and
$\chi_\rho,\ \chi_\lambda$ are the standard three quark spin and isospin 
wave-functions~\cite{hosaka_book,close_book}. By the use of these expressions,
the interaction Eq.~(\ref{eq:alt}) is reduced to
\begin{eqnarray}
{\cal
L}_{int}&=&2a(\cos\theta\phi_\rho\chi_\rho+\sin\theta\phi_\lambda\chi_\lambda)^\dagger(\cos\theta\phi_\rho\chi_\rho+\sin\theta\phi_\lambda\chi_\lambda)\nonumber\\
&+&2b(-\sin\theta\phi_\rho\chi_\rho+\cos\theta\phi_\lambda\chi_\lambda)^\dagger(
-\sin\theta\phi_\rho\chi_\rho+\cos\theta\phi_\lambda\chi_\lambda).
\label{eq:NR}
\end{eqnarray}
If we take $\tan\theta=1$,
we realize  SU(4)$_{SF}$ symmetry for the first term in
Eq.~(\ref{eq:NR}), while the second term become the SU(4)$_{SF}$
forbidden state. Since we employ the diquark correlations violating
SU(4)$_{SF}$ symmetry, the SU(4)$_{SF}$ forbidden state can be
included. Then, the two channel type interaction Eq.~(\ref{eq:twoc})
gives the two bound states of a quark and a diquark.

\end{document}